\documentclass[english,aps,prl,twocolumn]{revtex4}
\usepackage[T1]{fontenc}
\usepackage[latin1]{inputenc}
\usepackage{graphicx}
\usepackage{amssymb}

\begin{document}

\title{Intrinsic Quantum Noise in Faraday Rotation Measurements of a Single
Electron Spin}

\author{Yanjun Ma and Jeremy Levy}

\affiliation{Department of Physics and Astronomy, University of Pittsburgh, Pittsburgh, PA 15260}

\date{2 September 2008}

\begin{abstract}
Faraday rotation is one way to realize quantum non-demolition measurement
of electron spin in quantum dots. To describe Faraday rotation, semiclassical
models are typically used, based on quantized electron spin states
and classical electromagnetic fields. Such treatments neglect the
entanglement between electronic and photonic degrees of freedom that
produce intrinsic quantum noise, limiting the ultimate sensitivity
of this technique. We present a fully quantum-mechanical description
of Faraday rotation, and quantify this intrinsic noise. A method for
measuring the purity of a given spin state is suggested based on this
analysis.
\end{abstract}
\maketitle

\section{Introduction}

Because of the discovery of long-lived spin coherence in semiconductors such as GaAs\cite{kikkawa}, the essential requirement of manipulating spins for spintronics and quantum information is now possible. The first quantum computing
proposal of Loss and DiVincenzo used electron spin qubits in semiconductor
quantum dots (QD)\cite{divi} and forecast the importance of measuring
single electrons and their spins.

The first step to realizing coherent manipulation of a single electron
spin is to orient the spin. Such orientation can be achieved optically
(by exciting with circularly polarized light)\cite{opti}, electrically
(by driving the electrons toward a ferromagnetic surface)\cite{asw}
or thermodynamically (by application of a uniform magnetic field at
low temperatures). Photoluminescence (PL) allows
for measurement of electron spin polarization through the relation
between the circular polarization of light and electron spin orientation.
However, PL is destructive in that it involves recombination of the
electron with a hole. PL measurements are intrinsically limited by
the lifetime of the state, and it is not possible to monitor electron
spin continuously. Furthermore, unless one uses a technique such as
time-resolved upconversion\cite{upc1,upc2,upc3} or streak camera
measurements, dynamical information is lost.

Time-resolved Faraday and Kerr rotation methods (hereafter referred
to as Faraday rotation) have been extensively developed\cite{exp1,exp2,exp3},
and allow one to probe the spin dynamics of a single electron in a quantum
dot. Faraday rotation results from a fundamental interaction between
electronic and photonic degrees of freedom. Seigneur et al.\cite{spfe}have
proposed a scheme to implement quantum computation by using the single
photon Faraday effect. However, in most semiconductors the Faraday
effect is usually quite weak, corresponding to rotation angles $\theta_{F}\sim10^{-5}\mathrm{rad}$
for single electrons. Dynamic information is usually obtained using
pump-probe optical techniques: a circularly polarized pump beam creates
an initially spin-polarized electron population, and a probe beam subsequently
interrogates the spin state at a later time. The experiment is performed
repeatedly as a function of the delay to obtain a time-resolved signal
with an acceptably high signal-to-noise ratio. In the case of a single
electron in a quantum dot, spin coherence can be achieved in the following
manner: the quantum dot is configured (either through biasing or doping
) to begin in a state that contains a single electron in the conduction
band and no holes in the valence band. The quantum dot is excited,
promoting a second electron into the conduction band and leaving behind
a hole in the valence band. This state is often referred to as a {}``trion''.
After one of the electrons recombines with the hole, the remaining
electron spin is partially polarized. A linearly polarized probe pulse
measures the spin of this electron via the Faraday effect. In most
cases, the electron neither begins in a pure state nor remains in
one. Hyperfine interactions with nuclear spins quickly produce a mixed
state on time scales \textasciitilde{}1-10 ns\cite{nuc1,nuc2,nuc3,nuc4,nuc5,nuc6,nuc7}.
In this paper, we study the noise introduced by the mixed quantum state of the electron spin analytically and numerically. In this paper, our previous analysis\cite{patr} about the noise is extended to a more formal quantum mechanical frame. Since it's from the spin state itself, we call it intrinsic noise.

\begin{figure}
\centering \includegraphics[width=0.5\textwidth,height=0.3\textheight]{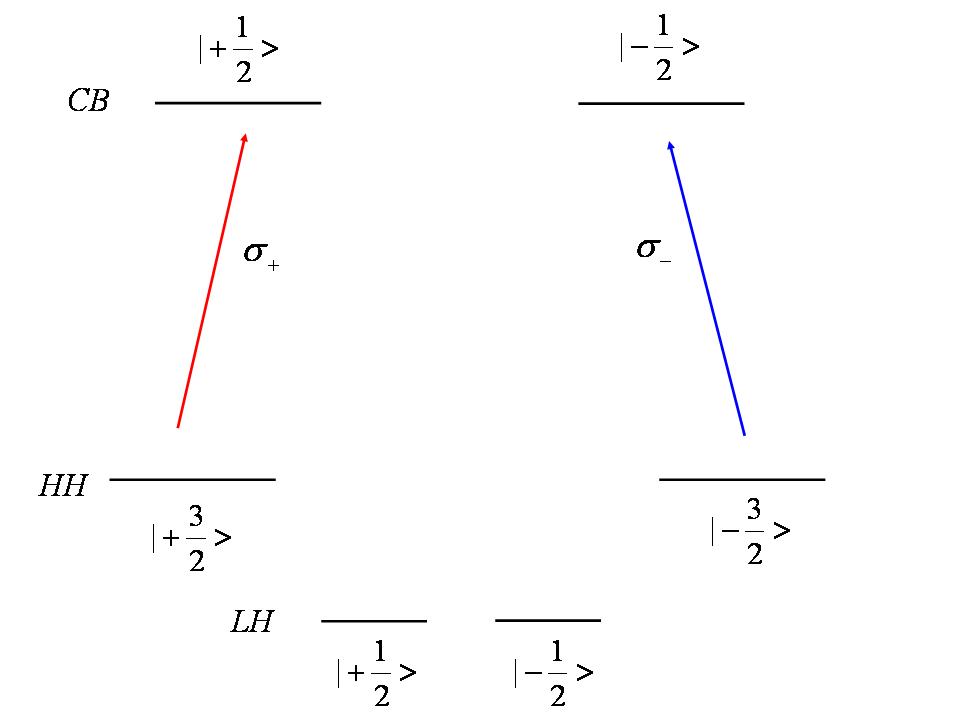}

\caption{light induced interband transition}

\label{band}
\end{figure}

\section{Theoretical Model}

Here we discuss in detail the quantum-mechanical source of this noise
using a theoretical model that treats both the electron and light
field quantum mechanically. We model the interaction between a single
electron in a QD and a linearly polarized monochromatic probe laser
field. The Hamiltonian for the photon field can be written as

\begin{equation}
H_{P}=\hbar\omega_{P}(a_{L}^{\dag}a_{L}+a_{R}^{\dag}a_{R}),
\end{equation}

 where $\omega_{P}$ is the optical frequency of the probe laser, $a_{L}^{\dag}$
and $a_{L}$ are creation and annihilation operators for left circularly
polarized (LCP) photons; $a_{R}^{\dag}$ and $a_{R}$ are creation
and annihilation operators for right circularly polarized (RCP) photons.
Due to optical selection rules\cite{opti} spin-up (spin-down) electrons
interact only with LCP (RCP) photons(See FIG.~\ref{band}). The raising
and lowering operators satisfy boson commutation relations \[
[a_{m},a_{n}^{\dag}]=\delta_{mn},(m,n=L,R);\]
 \[
[a_{m},a_{n}]=0,[a_{m}^{\dag},a_{n}^{\dag}]=0.\]

The electron state is quantized as well. We assume that the electron
resides in the conduction band quantum-confined ground state in an
s orbital, which means it has total angular momentum $J=\frac{1}{2}$.
In the valence band, the electronic ground states are constructed from
p-orbitals, and hence the total angular momentum is $J=\frac{3}{2}$.
The Hamiltonian for the electron is given by \cite{yama}

\begin{equation}
H_{e}=\hbar\omega_{e}(\sigma_{uz}+\sigma_{dz}),
\end{equation}

 where \[
\sigma_{uz}=b_{cu}^{\dag}b_{cu}-b_{vu}^{\dag}b_{vu},\]
 \[
\sigma_{dz}=b_{cd}^{\dag}b_{cd}-b_{vd}^{\dag}b_{vd};\]
 subscript \char`\"{}$c$\char`\"{} and \char`\"{}$v$\char`\"{} indicate
conduction band and valence band respectively; subscripts \char`\"{}$u$\char`\"{}
and \char`\"{}$d$\char`\"{} refer to spin-up or spin-down states
of the electron. The fermion operators satisfy anticommutation relations:
\[
\{b_{i\mu},b_{j\nu}^{\dag}\}=\delta_{ij}\delta_{\mu\nu},\]
 \[
\{b_{i\mu},b_{j\nu}\}=0,\{b_{i\mu}^{\dag},b_{j\nu}^{\dag}\}=0,\]
 where $i$ and $j$ indicate conduction band or valence band, and
$\mu$ and $\nu$ indicate spin-up or spin-down. Heavy-hole
and light-hole intermixing is neglected for simplicity and because it is not expected
to affect qualitatively our results. Only the heavy-hole subband is
accounted for in our calculation. A LCP photon couples to a transition
between $|+\frac{1}{2}>$ and $|+\frac{3}{2}>$, while a RCP photons
couples to a transition between $|-\frac{1}{2}>$ and $|-\frac{3}{2}>$.
The interaction Hamiltonian is given by

\begin{equation}
H_{I}=\lambda_{Lu}(a_{L}\sigma_{u+}+a_{L}^{\dag}\sigma_{u-})+\lambda_{Rd}(a_{R}\sigma_{d+}+a_{R}^{\dag}\sigma_{d-}),
\end{equation}

 where \[
\lambda_{Lu}\propto<+\frac{1}{2}|x+iy|+\frac{3}{2}>,\]
 \[
\lambda_{Rd}\propto<-\frac{1}{2}|x-iy|-\frac{3}{2}>,\]
 \[
\sigma_{u+}=b_{cu}^{\dag}b_{vu},\sigma_{u-}=\sigma_{u+}^{\dag},\]
 \[
\sigma_{d+}=b_{cd}^{\dag}b_{vd},\sigma_{d-}=\sigma_{d+}^{\dag}.\]
The full Hamiltonian of the entire system is given by \[
H=H_{P}+H_{e}+H_{I}\]
By applying the Wigner-Eckart theorem, it can be shown that the two
coupling strengths $\lambda_{Lu}$ and $\lambda_{Rd}$ must be equal
($\lambda_{Lu}$=$\lambda_{Rd}$$\equiv$$\lambda$). Based on the
defining anticommutation relations, it can be explicitly shown that
$\sigma_{\mu z}$,$\sigma_{\mu+}$ and $\sigma_{\mu-}$ have the following
commutation relations \[
[\sigma_{\mu+},\sigma_{\nu-}]=\delta_{\mu\nu}\sigma_{\mu z},\]
 \[
[\sigma_{\mu z},\sigma_{\nu+}]=2\delta_{\mu\nu}\sigma_{\mu+},\]
 \[
[\sigma_{\mu z},\sigma_{\nu-}]=2\delta_{\mu\nu}\sigma_{\mu-}.\]
These commutation relations for $\sigma_{\mu z}$,$\sigma_{\mu+}$
and $\sigma_{\mu-}$ are formally identical to those for the Pauli operators, even
though they are actually products of fermionic creation and annihilation
operators. This feature makes it possible to find an analytic solution
within the Heisenberg picture\cite{solution}. In the limit where
the coupling strength is much smaller than the incident photon frequency
or the characteristic frequency of the electron, the approximate solution
for photon operators is as follows:
\begin{equation}
a_{L}^{\dag}(t)=e^{-it\Omega\sigma_{uz}}a_{L}^{\dag}+g(t)(\sigma_{u+}+\alpha\sigma_{uz}a_{L}^{\dag}),\label{eq:al}
\end{equation}

\begin{equation}
a_{R}^{\dag}(t)=e^{-it\Omega\sigma_{dz}}a_{R}^{\dag}+g(t)(\sigma_{d+}+\alpha\sigma_{dz}a_{R}^{\dag}),\label{eq:ar}
\end{equation}

 where \[
\alpha=\frac{\lambda}{\omega_{P}-\omega_{e}},\]
 \[
\Omega=\lambda\alpha,\]
 \[
g(t)=\alpha(1-e^{-i(\omega_{P}-\omega_{e})t}).\]
This approximate solution is correct only when the coupling strength $\lambda$ is much smaller than $\omega_{e}$ and $\omega_{P}$. In the section of RESULTS, one can see this criteria is satisfied in the sense that the coupling strength of our sample is in the order of \textasciitilde{}$10^{9}$Hz, but the frequency of laser and characteristic frequency of electron is in the order of \textasciitilde{}$10^{15}$Hz. This solution, therefore, is a very good approximation and based on this, one can derive Faraday rotation angle.

\subsection{Faraday rotation operator}

\begin{figure}
\centering \includegraphics[width=0.5\textwidth,height=0.3\textheight]{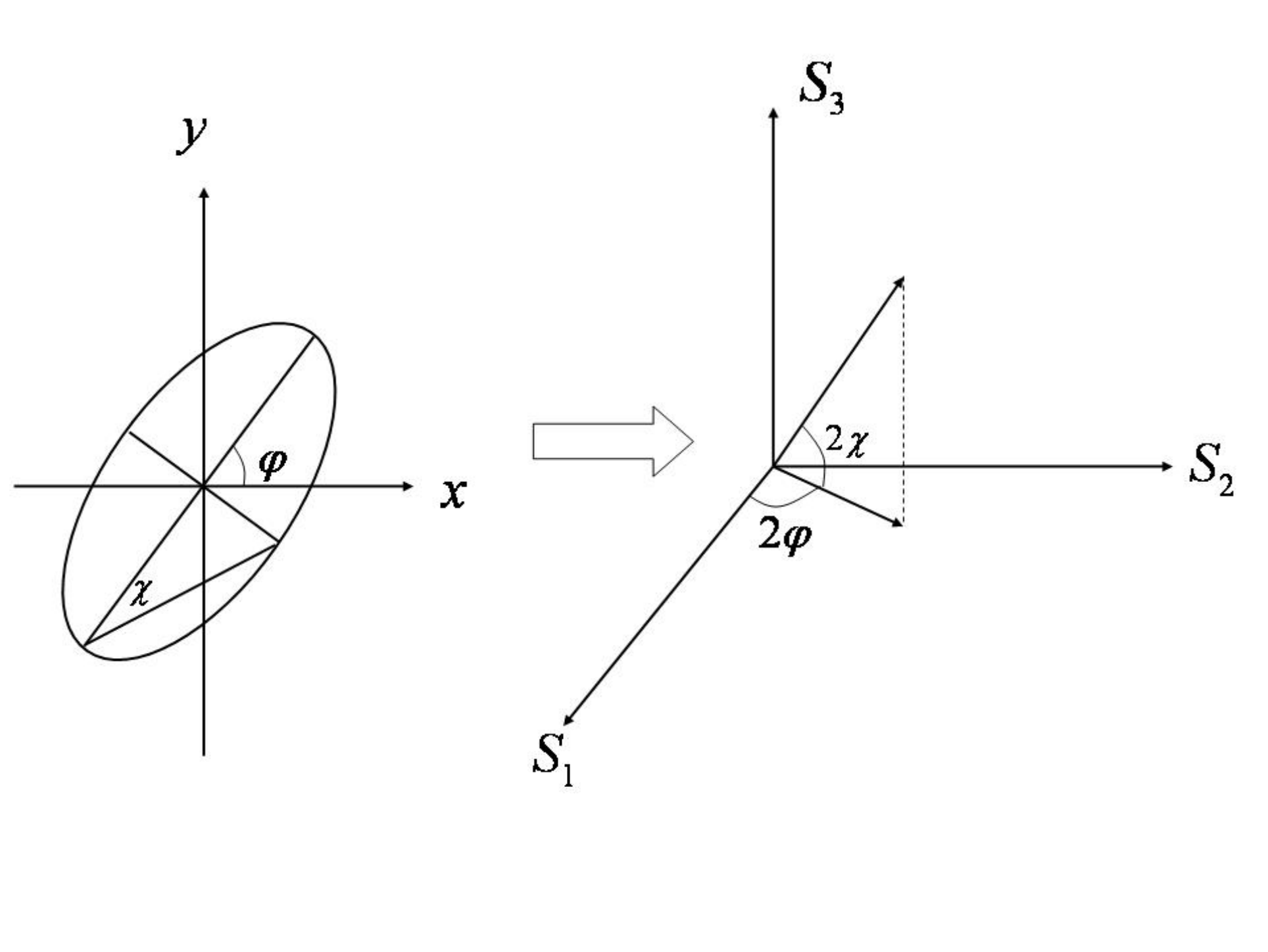}

\caption{The left figure shows the polarization ellipse in real space. The right figure is the Stokes representation of the same polarization.}

\label{stokerep}
\end{figure}

Quantum Stokes operators can be used to describe Faraday rotation.
They are the quantum-mechanical analogue of classical Stokes parameters.
Classical Stokes parameters are defined as the following\cite{stoke1}
\[
\left\{ \begin{array}{ll}
S_{0}=E_{x}^{*}E_{x}+E_{y}^{*}E_{y}\\
S_{1}=E_{x}^{*}E_{x}-E_{y}^{*}E_{y}\\
S_{2}=E_{x}^{*}E_{y}+E_{y}^{*}E_{x}\\
S_{3}=E_{x}^{*}E_{y}-E_{y}^{*}E_{x}\end{array}\right.\]
 In electrodynamics, the polarization of light can be parameterized
by two angles $\varphi$ and $\chi$ in the polarization ellipse. There
is a one-to-one correspondence between the polarization-ellipse representation
and the Stokes representation (See FIG.~\ref{stokerep}). %

\begin{figure}[h]
 \centering \includegraphics[width=0.5\textwidth,height=0.3\textheight]{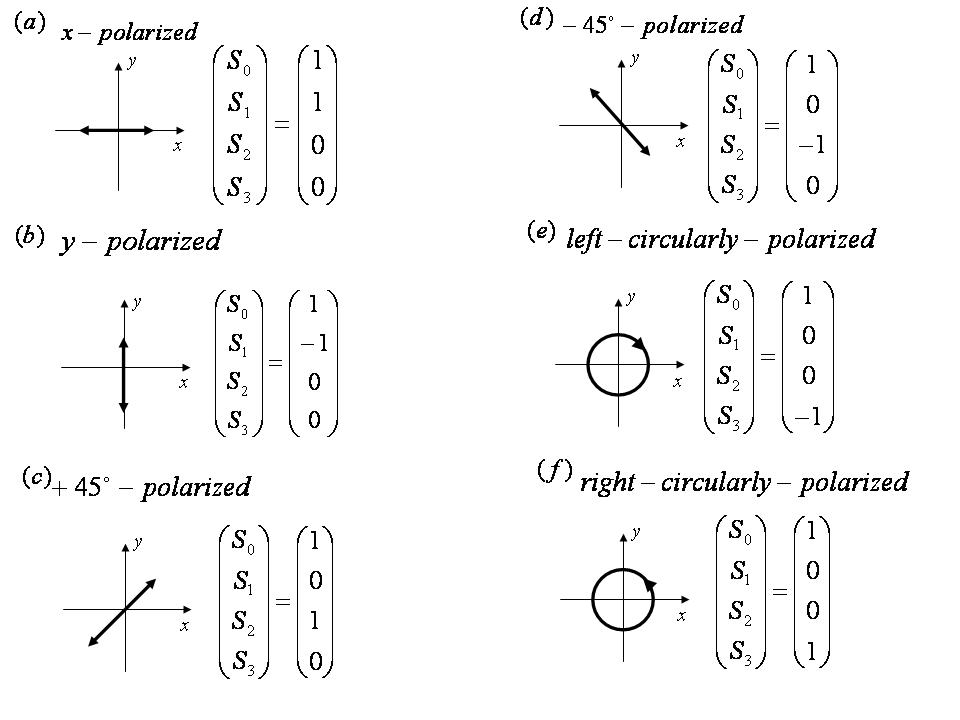}

\caption{(a)-(f)Different polarization defined in terms of Stokes parameters. $S_{1}, S_{2}$ and $S_{3}$ have all been normalized to $S_{0}$.}
\label{stokepolar}

\end{figure}

Once the light field is known, the Stokes parameters can be computed.
The physical interpretation of $S_{0}$ is the light intensity; hence,
all parameters can be normalized to $S_{0}$ (See FIG.~\ref{stokepolar}).

\begin{figure}
\centering \includegraphics[width=0.5\textwidth,height=0.3\textheight]{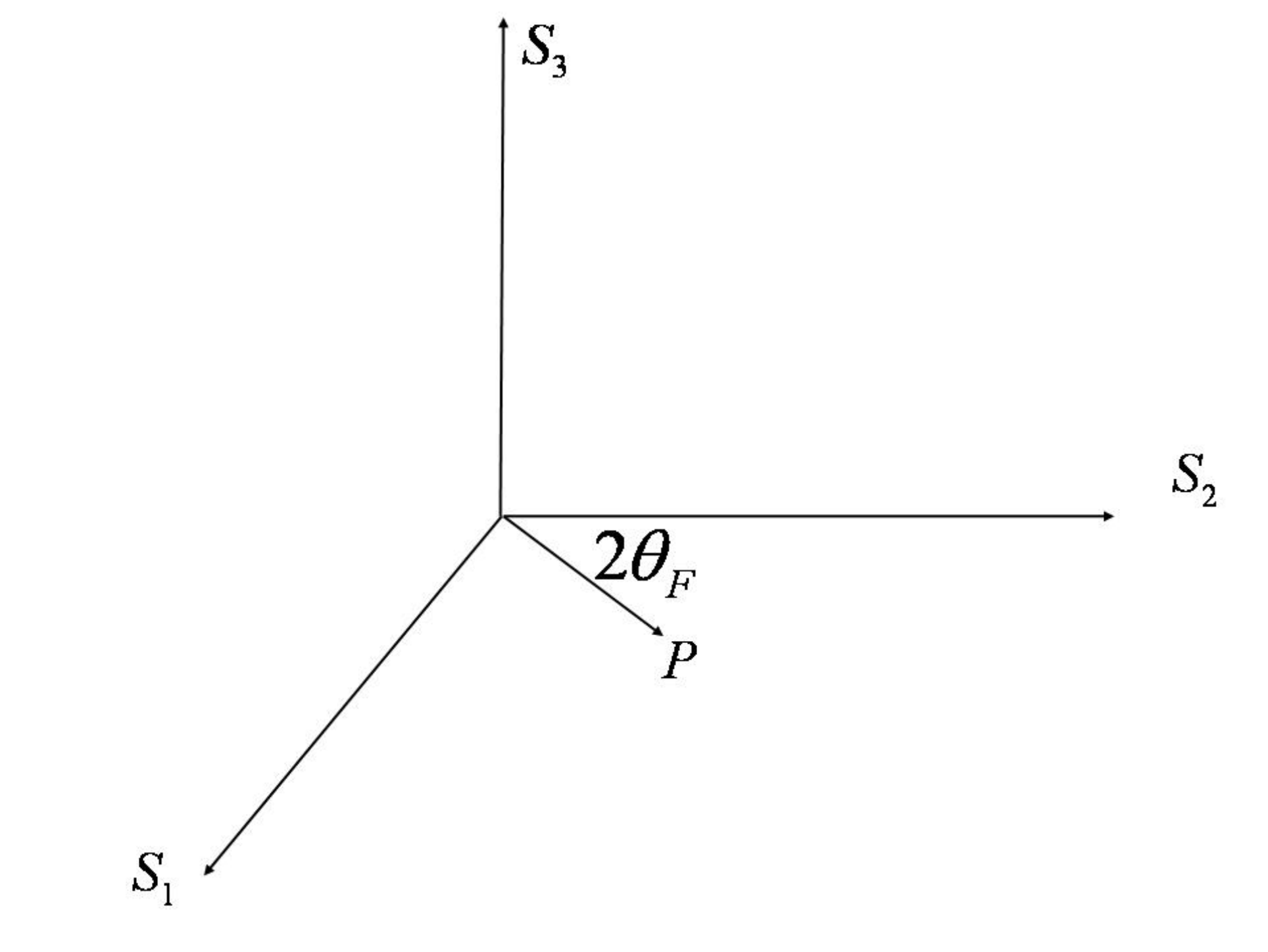}

\caption{Definition of Faraday rotation angle. P indicates the polarization vector of light field.}

\label{fangle}
\end{figure}

Quantum Stokes operators are defined in the following way\cite{stoke2,stoke3}
\[
\left\{ \begin{array}{l}
S_{0}=a_{L}^{\dag}a_{L}+a_{R}^{\dag}a_{R}\\
S_{1}=a_{L}^{\dag}a_{R}+a_{R}^{\dag}a_{L}\\
S_{2}=i(a_{L}^{\dag}a_{R}-a_{R}^{\dag}a_{L})\\
S_{3}=a_{R}^{\dag}a_{R}-a_{L}^{\dag}a_{L}\end{array}\right.\]
 Information about polarization is obtained by calculating the expectation
values of these operators. In a typical Faraday experiment, the probe
light is linearly polarized at a $45^{\circ}$ angle with respect to a final
polarizing beamsplitter. After the interaction between the probe light
and the electron, the polarization of the transmitted light will be rotated
from its initial position by an angle $\theta_{F}$, known as the
Faraday rotation angle. In the Stokes representation, the initial
polarization vector lies along the positive $S_{2}$ axis. Faraday
rotation will result in a rotation of the vector within the $S_{1}-S_{2}$
plane (See FIG.~\ref{fangle}). This vector P is confined to the plane
as long as there is no circular dichroism that can lead to a non-zero
expectation value for $S_{3}$.

In our calculations, we aim to reproduce the overall magnitude of
the rotation angle that has been reported in experimental work\cite{exp1, exp2, exp3}.
The experimentally observed rotation angle is small: $\theta_{F}\sim10^{-5}\mathrm{rad}$.
Hence, it can be expressed as
\begin{equation}
\theta_{F}=\frac{1}{2}tan^{-1}(\frac{<S_{1}>}{<S_{2}>})\approx\frac{<S_{1}>}{2<S_{2}>}.
\end{equation}

\section{Results}

In our calculation, a coherent state $|\nu_{L},\nu_{R}>$is used for
the light field, where $|\nu_{L}^{2}|$ and $|\nu_{R}^{2}|$ are the
average number of left and right circularly polarized photons. These
states satisfy the canonical eigenvalue equations for the (non-Hermitian)
photon annihilation operators: \[
a_{L}|\nu_{L},\nu_{R}>=\nu_{L}|\nu_{L},\nu_{R}>,\]
 \[
a_{R}|\nu_{L},\nu_{R}>=\nu_{R}|\nu_{L},\nu_{R}>,\]

Using the form $\nu_{L}=N_{L}e^{i\theta_{L}}$ and $\nu_{R}=N_{R}e^{i\theta_{R}}$,
the expectation value of Stokes operators in this coherent state can
be found
\begin{equation}
\left(\begin{array}{l}
<S_{0}>\\
<S_{1}>\\
<S_{2}>\\
<S_{3}>\end{array}\right)=\left(\begin{array}{l}
N_{L}^{2}+N_{R}^{2}\\
2N_{L}N_{R}cos(\theta_{L}-\theta_{R})\\
2N_{L}N_{R}sin(\theta_{L}-\theta_{R})\\
N_{R}^{2}-N_{L}^{2}\end{array}\right).
\end{equation}
 In order to start with $+45^{\circ}$ linearly polarized light, the
following condition must be satisfied \[
\left\{ \begin{array}{l}
N_{L}^{2}=N_{R}^{2}\\
\theta_{L}-\theta_{R}=\frac{\pi}{2}\end{array}\right.\]

To describe the mixed state of electron, a density matrix formula
is employed. \begin{equation}
\rho_{e}=\tau|+\frac{1}{2}><+\frac{1}{2}|+(1-\tau)|-\frac{1}{2}><-\frac{1}{2}|\end{equation}

Here, $\tau$ is a parameter that varies between 0 and 1. For $\tau=0$
and $\tau=1$, one has a pure state, while $\tau=1/2$ corresponds
to a fully mixed (unpolarized) state. Because the electron Hamiltonian
is expressed in terms of creation and annihilation operators, caution
must be taken when applying those operators onto electron state. When
operators for spin-up electron are applied to the spin-up state, one
obtains \[
\sigma_{uz}|+\frac{1}{2}>=|+\frac{1}{2}>,\]
 \[
\sigma_{uz}|+\frac{3}{2}>=-|+\frac{3}{2}>,\]
 and \[
\sigma_{u+}|+\frac{1}{2}>=0,\]
 \[
\sigma_{u+}|+\frac{3}{2}>=|+\frac{1}{2}>,\]
 and \[
\sigma_{u-}|+\frac{1}{2}>=|+\frac{3}{2}>,\]
 \[
\sigma_{u-}|+\frac{3}{2}>=0.\]

Spin-down operators have the same rules when applied to the spin-down
state. If a spin-up operator operates on a spin-down state, however,
one gets zero. For example, \[
\sigma_{uz}|-\frac{1}{2}>=(b_{cu}^{\dag}b_{cu}-b_{vu}^{\dag}b_{vu})|-\frac{1}{2}>=0.\]

The initial state of the whole system is then \begin{equation}
\rho_{0}(\tau)=|\nu_{L},\nu_{R}><\nu_{L},\nu_{R}|\otimes(\tau|\uparrow_{-}>+(1-\tau)|\downarrow_{-}>)\end{equation}

According to the solution (4) and (5), the analytical expression for
$S_{1}$ and $S_{2}$ can be obtained and the expectation values calculated
\[
<S_{1}>=Tr(S_{1}(t)\rho_{0}(\tau)),\]
\[
<S_{2}>=Tr(S_{1}(t)\rho_{0}(\tau)).\]

The rotation angle is given by
\begin{equation}
\theta_{F}(t,\tau)=\frac{Tr(S_{1}(t)\rho_{0}(\tau))}{2Tr(S_{2}(t)\rho_{0}(\tau))},
\end{equation}
 After some algebra, one finds the following expression for the Faraday
rotation:
\begin{equation}
\theta_{F}(t,\tau)=(2\tau-1)(\frac{\lambda^{2}}{\delta^{2}}sin(\delta t)-sin(\frac{\lambda^{2}}{\delta}t)),
\end{equation}
 where $\delta\equiv\omega_{P}-\omega_{e}.$ For initial pure spin-up
state $\tau=1$, the rotation angle is
\begin{equation}
\theta_{+}\equiv\theta_{F}(t,1)=(\frac{\lambda^{2}}{\delta^{2}}sin(\delta t)-sin(\frac{\lambda^{2}}{\delta}t)).
\end{equation}
 For initial pure spin-down state $\tau=0$, the rotation angle is
\begin{equation}
\theta_{-}\equiv\theta_{F}(t,0)=-(\frac{\lambda^{2}}{\delta^{2}}sin(\delta t)-sin(\frac{\lambda^{2}}{\delta}t)).
\end{equation}

The fluctuation is given by
\begin{equation}
\Delta\theta_{F}(t,\tau)=\frac{\sqrt{Tr(S_{1}^{2}(t)\rho_{0}(\tau))-Tr(S_{1}(t)\rho_{0}(\tau))^{2}}}{2Tr(S_{2}(t)\rho_{0}(\tau))}.
\end{equation}

\begin{figure}
\centering \includegraphics[width=0.5\textwidth,height=0.3\textheight]{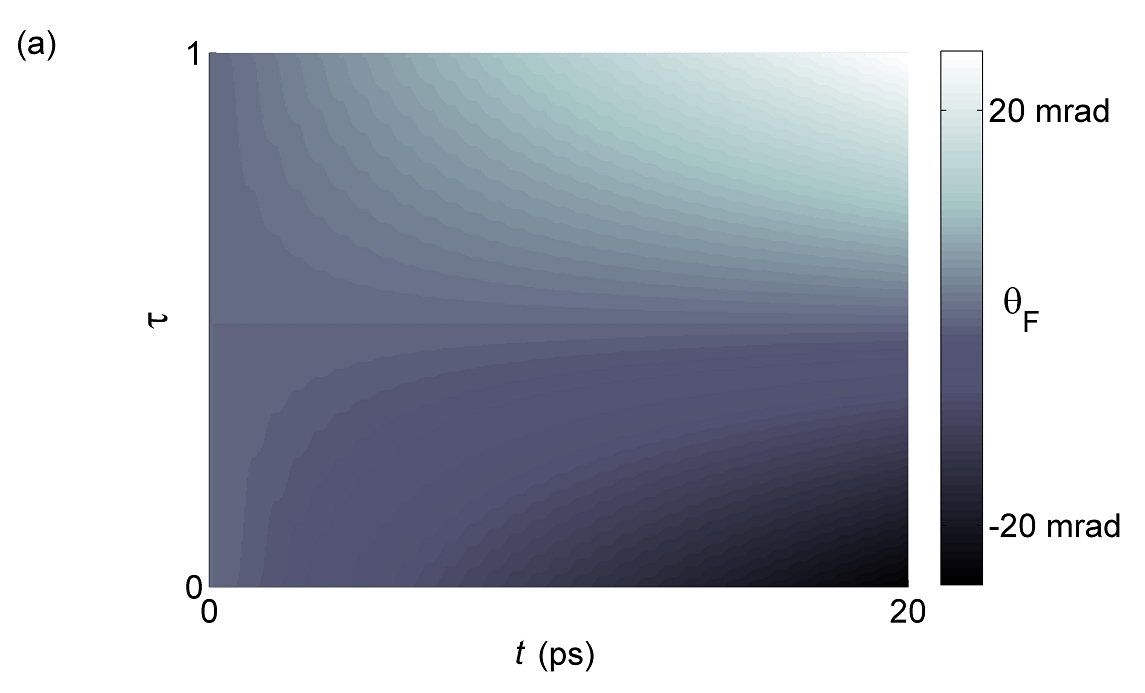}
\hfill{}\includegraphics[width=0.5\textwidth,height=0.3\textheight]{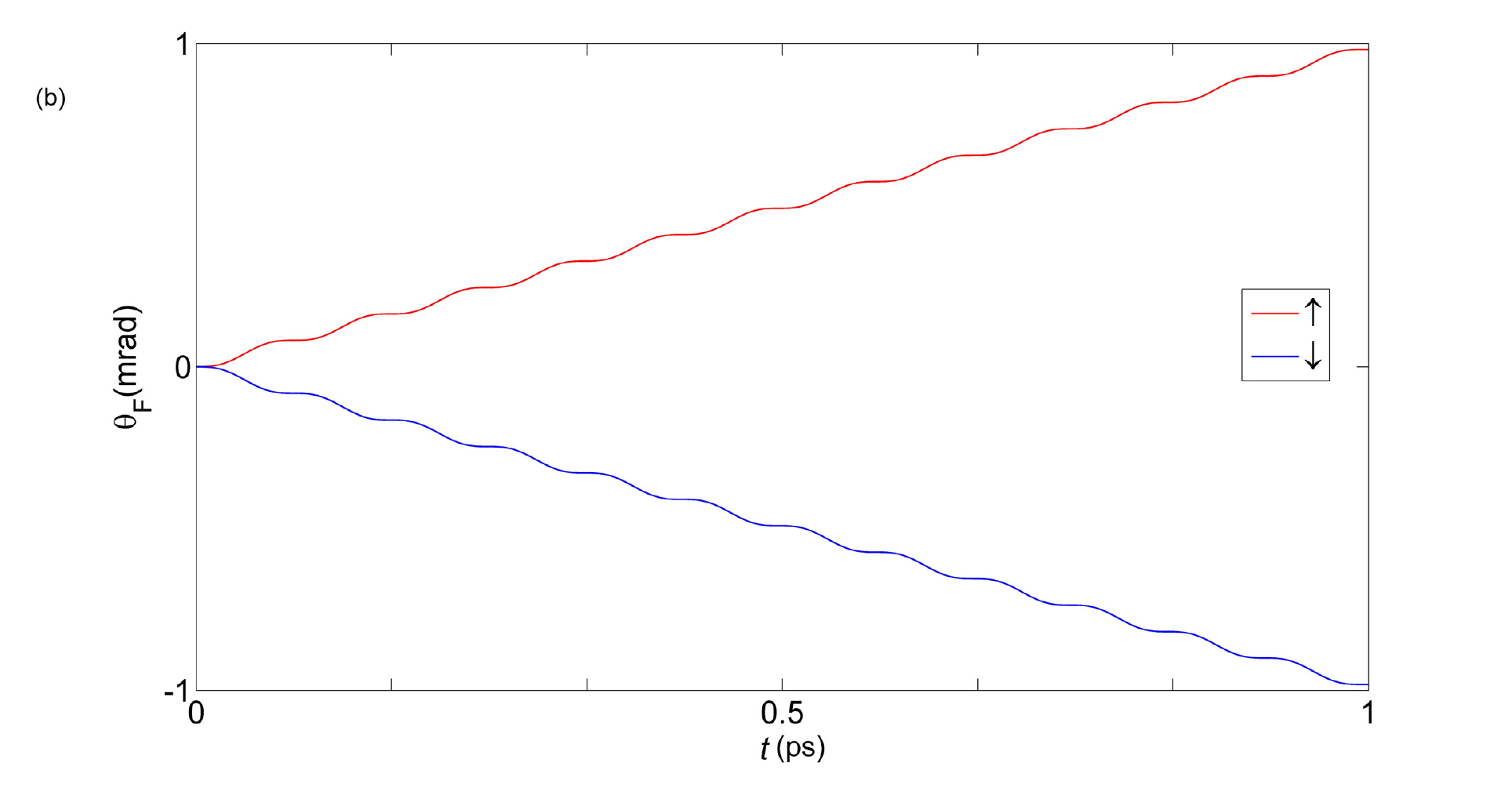}

\caption{(a)Faraday rotation as a function of time and parameter $\tau$. (b)Faraday
rotation angle for two pure states. $\tau=1$ corresponds to a spin-up
state, while $\tau=0$ corresponds to a spin-down state. }

\label{angle}
\end{figure}

From (11), (12) and (13), the following intuitive result can be proven
very easily: \begin{equation}
\theta_{F}(t,\tau)=\tau\theta_{+}+(1-\tau)\theta_{-},\end{equation}

where $\theta_{+}$ ($\theta_{-}$) is the Faraday rotation angle
for an initial state which is a pure spin-up (spin-down) state.

An analytical derivation shows that the fluctuation is a function
of both photon number and the initial electron state.
\begin{equation}
\Delta\theta_{F}(t,\tau)=\sqrt{\frac{1}{4N}+\tau(1-\tau)(\theta_{+}-\theta_{-})^{2}},
\end{equation}
 The second term under the square root is the so-called intrinsic noise term.

\begin{figure}
\centering
\includegraphics[width=0.5\textwidth,height=0.3\textheight]{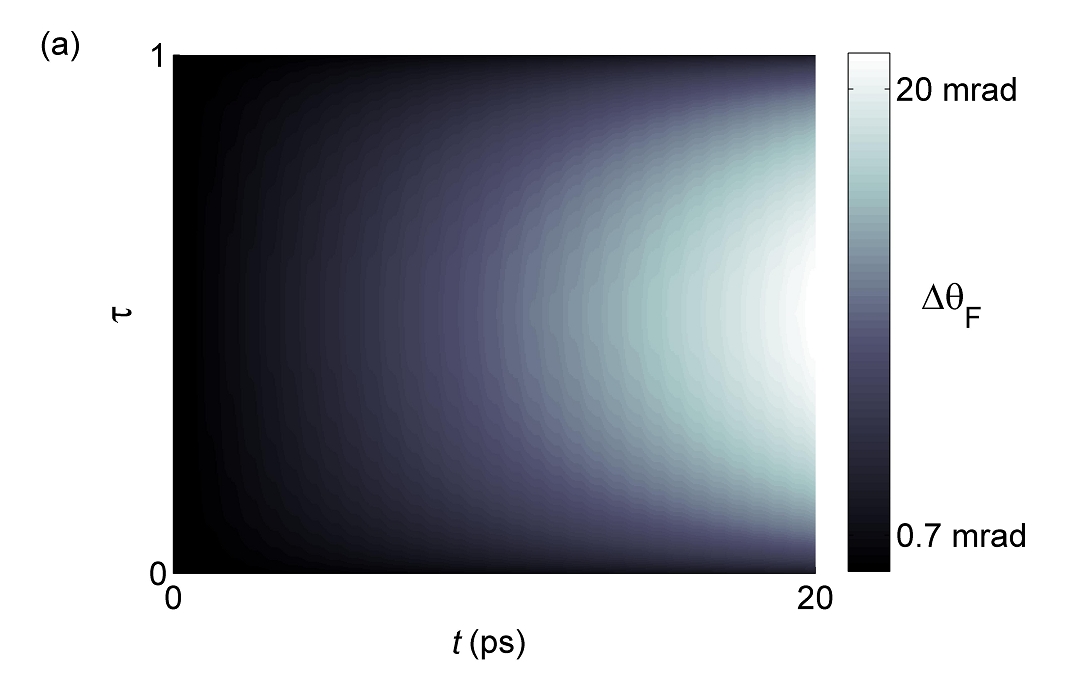}
\hfill{}
\includegraphics[width=0.5\textwidth,height=0.3\textheight]{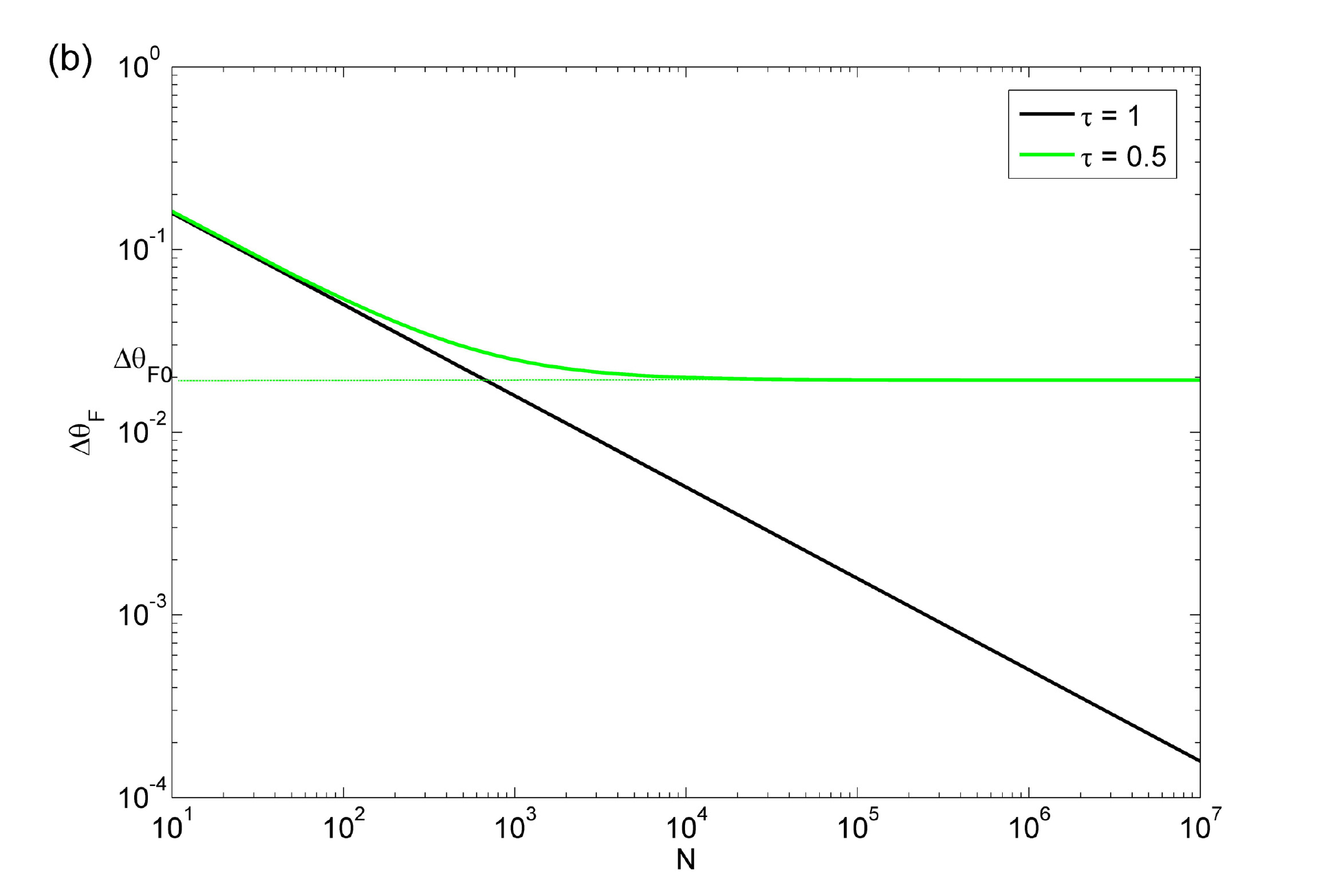}

\caption{(a)Fluctuation of Faraday rotation angle as a function of time and parameter $\tau$. (b)Shot noise and intrinsic noise as a function of photon number N.Shot noise (black) is from a pure spin-up(spin-down) state, while intrinsic noise (green) is from a maximally mixed state. In the simulation, the interaction time is chosen to be $20\mathrm{ps}$ in order to make the splitting more obvious, in which case the intrinsic noise saturates at about $\Delta\theta_{F0}=20mrad$.}

\label{noise}
\end{figure}

Numerical simulation is done so that we compare our analytical calculation to recent Kerr rotation experimental results on single electrons. From Berezosky et al.\cite{exp1}, one finds from
a PL plot that the energy for a neutral exciton is about $1.633\mathrm{meV}$.
That corresponds to the band gap between the top of the valence band and
the bottom of the conduction band. From this number, the frequency $\omega_{e}=\frac{E}{\hbar}=2.48\times10^{15}Hz$.
Choosing probe light of wavelength $760\mathrm{nm}$, which means the
frequency is $\omega_{P}=2.47\times10^{15}Hz$. From the paper\cite{cpl},
one can take the value of the coupling strength to be $\lambda=\mathrm{98GHz}$. In our experiment, the probe power is about $1.57\mathrm{\mu W}$, the corresponding photon number is about $5\times10^{5}$. In the simulation, the interaction time between the spin and the photon is set to be $20\mathrm{ps}$. Notice that as expected, if the initial electron state is a pure state, the rotation angle has opposite values for the spin-up state and spin-down state,
respectively (See FIG.~\ref{angle}). For pure spin-up states or
spin-down states, the fluctuation (quantum noise) scales with photon
number $N$ as $N^{-1/2}$, as expected for shot noise. For mixed
states or superposition states, the fluctuation saturates even when
the photon number approaches infinity (See FIG.~\ref{noise}).

One scheme to measure $\tau$ is proposed here. Suppose the photon number is so large that the shot noise term in Equation (16) could be neglected. Notice that $\theta_{+}=\theta_{-}$ and when the rotation
angle is zero, according to Equation (11) it means $\tau$ is $\frac{1}{2}$.
If the value $\tau=\frac{1}{2}$ is used in Equation (16), one obtains
$\Delta\theta_{F0}=\theta_{+}$. This result implies that one can
use the measured values of Faraday angle fluctuation at an extreme
value ($\Delta\theta_{F}$) and at a zero crossing ($\Delta\theta_{F0}$)
to measure the purity of the spin state as quantified by $\tau$:
\begin{equation}
\tau=\frac{1}{2}(1\pm\sqrt{\frac{\Delta\theta_{F0}^{2}-\Delta\theta_{F}^{2}}{\Delta\theta_{F0}^{2}-\frac{1}{4N}}}).
\end{equation}

In the limit of a large number of photons (i.e., where shot noise can
be neglected), the above expression simplifies further
\begin{equation}
\tau=\frac{1}{2}(1\pm\sqrt{1-\frac{\Delta\theta_{F}^{2}}{\Delta\theta_{F0}^{2}}}).
\end{equation}

The above analysis is based on the assumption that every device in
the experiment is perfect, and the noise is only introduced by quantum
state of the electron spin itself. This is, however, not the case in the
real experiment. Suppose the overall noise $\Delta\theta_{B}$ is
white noise for the bandwidth in which the experiment is done. It
serves as background noise and can be measured by detuning the probe
laser, for example. This background contribution can be subtracted
from the measured noise $\Delta\theta_{M}$ and $\Delta\theta_{M0}$,
where $\Delta\theta_{M}$ indicates the measured noise level at extreme
points while $\Delta\theta_{M0}$ represents the measured noise at
the zero-crossing point. It makes sense to assume the fluctuation due
to quantum states is not correlated with the white noise in the device,
therefore subtracting the background noise from the measured noise
gives the fluctuation due to quantum states \[
\Delta\theta_{F}^{2}=\Delta\theta_{M}^{2}-\Delta\theta_{B}^{2},\]
 and \[
\Delta\theta_{F0}^{2}=\Delta\theta_{M0}^{2}-\Delta\theta_{B}^{2}.\]
 The above equation therefore becomes \begin{equation}
\tau=\frac{1}{2}(1\pm\sqrt{1-\frac{\Delta\theta_{M}^{2}-\Delta\theta_{B}^{2}}{\Delta\theta_{M0}^{2}-\Delta\theta_{B}^{2}}}).\end{equation}

Notice that in the above equation, $\Delta\theta_{M}$ is the \char`\"{}real\char`\"{}
noise we see at the extreme points of the rotation angle in the actual
experiment. This noise has two sources: external noise, which is $\Delta\theta_{B}$
and intrinsic noise, which is $\Delta\theta_{F}$. In the actual experiment,
$\Delta\theta_{M}$ should be larger than $\Delta\theta_{B}$ due
to the fact that the pump for the electron spin is not perfect, therefore
the spin that interacts with photon is in a mixed state. However,
if $\Delta\theta_{M}=\Delta\theta_{B}$, that means the intrinsic
noise contribution is zero. From Eq. (16), one can see in the limit
of large photon number, $\Delta\theta_{F}=0$ indicates that $\tau$
is either 0 or 1, which is consistent with the result if one plugs
$\Delta\theta_{M}=\Delta\theta_{B}$ into Eq. (19). In other words,
if in the real experiment, one observes $\Delta\theta_{M}=\Delta\theta_{B}$,
then the pumped spin is in either pure spin-up or spin-down state
and one can also pin down the orientation of spin by looking at the
sign of measured rotation angle.

\section{Conclusion}

Using a quantum-mechanical model of Faraday rotation, we find that
both the Faraday rotation angle and the fluctuation are functions of the initial
electron spin state. If the electron spin is initially in a mixed
state, intrinsic noise fluctuations will contain not only shot noise
but also intrinsic noise due to weak measurement of the electron's
spin state. The reason that this intrinsic noise appears in this scheme
is that the measurement done here is non-destructive, and differs
from a projective measurement, which causes the collapse of the electron
spin wave function to a certain spin direction. Analysis of the noise
spectrum should enable quantification of the purity of a given spin
state.

\end{document}